\documentclass[twocolumn,showpacs,amsmath,amssymb,superscriptaddress]{revtex4-1}
\usepackage{dcolumn,braket,color,amsmath,footnote,hyperref,tabularx,multirow,graphicx,bm,url,longtable,verbatim,epic,eepic,babel,booktabs,feynmp,epsfig,xcolor,hyperref}
\usepackage[caption = false]{subfig}
\def\orcid#1{\kern .08em\href{https://orcid.org/#1}{\includegraphics[keepaspectratio,width=0.7em]{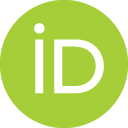}}} 
\begin{document}
\pagestyle{plain}
\title{Transfer reactions between odd-odd and even-even nuclei by using IBFFM}

\author{Ruslan Maga\~na Vsevolodovna 
\orcid{0000-0001-6754-4238} 
}
\email{ruslan.magana@ge.infn.it}
\affiliation{INFN, Sezione di Genova and Universit\`a di Genova, via Dodecaneso 33, 16146 Genova, Italy}

\author{Elena Santopinto
\orcid{0000-0003-3942-6554} 
}
\email{corresponding author:santopinto@ge.infn.it}
\affiliation{INFN, Sezione di Genova and Universit\`a di Genova, via Dodecaneso 33, 16146 Genova, Italy}

\author{Roelof Bijker \orcid{0000-0002-7266-934X} 
\\for the NUMEN collaboration}
\email{bijker@nucleares.unam.mx}
\affiliation{Instituto de Ciencias Nucleares, Universidad Nacional Aut\'onoma de M\'exico, A.P. 70-543, 04510 M\'exico D.F., Mexico}

\date{\today}

\begin{abstract}
Spectroscopic Amplitudes (SA) in the Interacting Boson Fermion Fermion Model (IBFFM) are necessary for the computation of $0\nu\beta\beta$ decays but also for cross sections of heavy-ion reactions, in particular, Double Charge Exchange reactions for the NUMEN collaboration, if one does not want to use the closure limit. 
We present for the first time: i) the formalism and operators to compute in a general case the spectroscopic amplitudes in the scheme IBFFM from an even-even to odd-odd nuclei, in a way suited to be used in reaction code, i.e., extracting the contribution of each orbital; 2) the odd-odd nuclei as described by the old IBFFM are obtained for the first time with the new implementation of Machine Learning (ML) techniques for fitting the
parameters, getting a more realistic description. 
The one body transition densities for $^{116}$Cd $\rightarrow$ $^{116}$In and $^{116}$In $\rightarrow$ $^{116}$Sn are part of the experimental program of the NUMEN experiment, which aims to find constraints on Neutrinoless double beta decay matrix elements.
\end{abstract}


\maketitle


The first articles on odd-odd nuclei with algebraic models date back to the 80s when some odd-odd nuclei were studied by dynamical symmetry scheme, and a supersymmetry scheme \cite{Isacker1985,Warner1986,Balantekin1986}.
Unfortunately, that approach is suitable only for a few nuclei that correspond to dynamical supersymmetry. 
It must be emphasized that nuclear supersymmetry places very severe constraints on the form of the boson-fermion hamiltonian used to describe nuclei belonging to a supermultiplet. Consequently, applications are restricted to only a few regions of the nuclear chart.
The first possible example of a supermultiplet involving an odd-odd nucleus was proposed in the Pt-Au region \cite{Isacker1985,Chou1990} 
and subsequent investigations \cite{Balantekin1986,Warner1986} focused on the odd-odd member of the quartet, $^{198}$Au in \cite{Mayerhofer1989}, $^{196}$Au in \cite{Metz1999,Barea_2004,Barea2005} and $^{194}$Ir in \cite{Balodis2008,Barea2009-2}. 

In a second more general approach, which we will follow in this paper, 
the Interacting Boson Model (IBM) was extended to describe odd-odd nuclei, 
in which the two fermions (one neutron and one proton) are coupled to the even-even core described either by IBM-1 \cite{Lopac1986,Blasi1987,Brant1988}, 
or by IBM-2 \cite{Yoshida2013,Nomura2020}. These extensions to odd-odd nuclei 
are usually referred to as the Interacting Boson-Fermion-Fermion Model, IBFFM-1 
and IBFFM-2, respectively. In this article we use the neutron-proton version 
which we will use the shorthand notation IBFFM from here on (strictly speaking it is IBFFM-2). 
 
The IBFFM hamiltonian is derived from semi-microscopic arguments and subsequently diagonalized numerically. It has been introduced by Lopac and Bianco \cite{Lopac1986,Blasi1987}, and it is essentially an IBFM-1 in the proton, and an IBFM-1 in the neutron plus a residual interaction as introduced by Brant and Paar \cite{Brant1988}. The original approach was extended to include the neutron-proton 
degree of freedom in the description of the even-even core nucleus in Ref.~\cite{Yoshida2013}. In this reference Yoshida and Iachello and Yoshida wrote the operator expression for single beta decay and two neutrino double beta decay, 
which can be considered as a particular subset of the general one-body transition operator presented in this paper. 

The aim of this paper is from one side to investigate the transition from $^{116}$Cd to $^{116}$In,  and  $^{116}$Sn to $^{116}$In and introduce the formalism of the Spectroscopic Amplitudes (SA) between even-even and odd-odd nuclei in the IBFFM that are the key ingredient 
for planned publications of Double Charge Exchange (DCE) reactions for the NUMEN collaboration and the study  0$\nu\beta\beta$  decay.

The spectroscopic amplitudes operator dressed as a charge exchange or as simple transition amplitudes is needed for the theoretical NUMEN 
project Ref.~\cite{Cappuzzello2018,Agodi2015} to be inserted into a reaction code. 

For the NUMEN experiment at INFN-LNS in particular there is interest for $^{116}$Cd $\rightarrow$ $^{116}$In and $^{116}$In $\rightarrow$ $^{116}$Sn
~\cite{PhysRevC.102.044606}.

In this article, the SA operator in the scheme of IBFFM has been derived in a general way and then calculated for the previous reactions and the results given in the form of spectroscopic amplitudes, which later on can be applied to double charge exchange, 
$0\nu\beta\beta$ decay processes without closure approximation. 
We introduce a new method to fit the odd-odd nuclei parameters by creating a multidimensional theoretical dataset using machine learning libraries such as Scikit-learn. 
The operators needed to compute the transitions between odd-odd and even-even nuclei by using IBFFM depends on two factors: 
the similarity between the states in the initial and final nucleus, which differ by two nucleons, and the transferred pair of nucleons' correlation.  
We have derived the two different transition operators' cases, the case where the numbers of bosons are conserved between
the even-even and odd-odd nuclei and the one where is not conserved, respectively. 

The organization of this paper is the following:
In Sec.~\ref{sec2} we briefly review a simple theory of the one body transitions density.  
In Sec.~\ref{sec3} we present the derivation of the transition operator for the IBFFM.
In Sec.~\ref{sec4} we discuss the theory of the odd-odd nucleus in the implemented IBFFM with Machine Learning (ML).  
In Sec.~\ref{sec5} we compute the nuclear wave functions and spectroscopic amplitudes for the transitions $^{116}$Sn to $^{116}$In and $^{116}$Cd to $^{116}$Sn.
In our last Sec.~\ref{sec6} we present our conclusions and discuss our results, also we give the future outlook.


\section{Spectroscopic amplitudes\label{sec2}}


The microscopic theory for direct charge exchange process by heavy ions following the approach by Greiner \cite{Kim1979}, and Etchogoyen \cite{Etchegoyen1989} 
requires the One Body Transition Densities (OBTD) that are needed in the form factors of the direct charge exchange reactions.
Nuclear initial and final states can be presented in the proton-neutron formalism, 
the one body transition densities without the isospin indices can be represented by
\begin{equation}
\mbox{OBTD}(A B;\lambda )=\frac{\langle J_{B}\| [a_{\rho}^{\dagger}\times\tilde a_{\rho'}]^{(\lambda )}\|J_{A}\rangle }{\sqrt{(2\lambda+1)}},
\end{equation}
where $|J_{A}\rangle$ represents the vector of the state of the initial nucleus
and $|J_{B}\rangle$ the final nucleus, 
the subindex $\rho (\rho')$ refers to a proton(neutron), the tilde denotes a time-conjugated state 
$\tilde a_{\rho_{j}}=(-1)^{j-m}a_{j,-m}$. We define the Spectroscopic 
Amplitudes (SA) as the the reduced matrix element
$\langle J_{B}\| [a_{\rho}^{\dagger}\times\tilde a_{\rho'}]^{(\lambda )}\|J_{A}\rangle $.

In the isospin formalism, the nuclear states have certain isospin, which results in a non-trivial isospin factor.
The IBFFM can be mapped into the proton-neutron formalism allowing to derive the one body transition densities for the bosonic-fermionic space.
In the followings sections, we present the derivation of the OBTD operator for the IBFFM framework, and we study the transitions $^{116}$Sn to the $^{116}$In, and $^{116}$In to the $^{116}$Cd that are of interest for the NUMEN experiment.


\section{Transition operator \label{sec3} }


The one body transition density operator 
from even-even to odd-odd in the IBFFM formalism can be obtained considering the 
mapping from the fermionic space into the boson-fermion-fermion space of the tensor operator of order $\lambda$ given by

\begin{equation}\begin{array}{lll}
[a_{\rho}^{\dagger}\times\tilde a_{\rho'}]^{(\lambda )}
\rightarrow
[P^{\dagger}_{j_\rho} \times \tilde P_{j'_{\rho}}]^{(\lambda)} \equiv {T}_{j_{\rho}j'_{\rho}}^{(\lambda)} ,
\end{array}
\label{mapping}
\end{equation}

where $a^{\dagger }_{j_{\rho}}$ and $\tilde a_{j_{\rho'}}$ are the fermion creation and annihilation operators,
while $P^{\dagger}_{j_\rho}$ and $ \tilde P_{j'_\rho}$ 
are the single nucleon transfer operator in the IBFM \cite{Scholten1985} scheme.

The operator for one nucleon transfer which the number of bosons is conserved is given as 
\begin{equation}
P^{\dagger}_{j_{\rho}}=\xi_{j_{\rho}}a_{j_{\rho}}^{\dagger}+\displaystyle \sum_{j'_{\rho}}\xi_{j_{\rho}j'_{\rho}}[[s^{\dagger} \times \tilde d_{j_{\rho}}]^{(2)}\times a_{j'_{\rho}}^{\dagger}]^{(j_{\rho})},
\label{transfer1a} 
\end{equation}

where $s^{\dagger}$ is the creation operator, 
and   $s=\tilde s$    annihilation operator  of  $s$-boson respectively,
and $\tilde d$ is related 
to the $d$-boson annihilation operator by $\tilde d_\mu= (-1)^{\mu} d_{-\mu}$.

In case the number of bosons is changed by one unit 
\begin{equation}
P^{\dagger}_{j_{\rho}}=\theta_{j_{\rho}}(s^{\dagger} \times \tilde a_{j_{\rho}})^{(j_{\rho})}+\displaystyle \sum_{j'_{\rho}}\theta_{j_{\rho}j'_{\rho}}[d^{\dagger} \times \tilde a_{j'_{\rho}}]^{(j_{\rho})},
\label{transfer1b} 
\end{equation}
where the subindex refers to a proton or a neutron occupying the orbitals $j_\pi$ and $j_\nu$, respectively. The coefficients $\xi_{j_{\rho}}$, $\xi_{j_{\rho}j'_{\rho}}$ and $\theta_{j_{\rho}}$, $\theta_{j_{\rho}j'_{\rho}}$ 
are the particle- or hole-coupling coefficients defined in Ref.~\cite{Scholten1985,Iachello1991}. 

The transitions between even-even nuclei and odd-odd nuclei can be computed by considering the tensorial product of the transfer operator 
of a particle and a hole coupled to the angular momenta $\lambda$, which is the value of the spin of the final state of the odd-odd nucleus.
The case where the number of bosons $N_{\nu},N_{\pi}$ between the even-even nucleus and 
the odd-odd core nucleus does not change, and the core of the odd-odd nucleus has attached a proton and removed a neutron
the operator is written as
\begin{eqnarray}
&& {T}_{ j_{\nu}j_{\pi}}^{(\lambda)} = 
\xi_{ j_{ \nu}} \xi_{j_{\pi}}
[
\tilde a_{ j_{\nu}} \times
   a_{j_{\pi}}^{\dagger}
 ]^{(\lambda)}
\nonumber\\
&& \quad + \xi_{j_{\nu}}
\displaystyle \sum_{j_{\pi}'}
\xi_{j_{\pi}j_{\pi}'}
[
\tilde a_{j_{\nu}}\times
[(s_{\pi}^{\dagger} \times \tilde 
d_{\pi})^{(2)}\times 
 a^{\dagger}_{j_{\pi}'}]^{ (j_{\pi})}
]^{(\lambda)}
\nonumber\\
&& \quad + \xi_{j_{\pi}} 
 \displaystyle \sum_{ j_{\nu}'}
\xi_{ j_{\nu}j_{\nu}'}
[[\tilde a_{j_{\nu}'} \times ( d_{\nu}^{\dagger} \times \tilde s_{\nu} )^{(2)}]^{(j_{\nu})}
\times   a_{j_{\pi}}^{\dagger}
 ]^{(\lambda)}
\nonumber\\
&& \quad + 
\displaystyle \sum_{j_{\nu}',j_{\pi}'}
\xi_{j_{\nu}j_{\nu}'}
\xi_{j_{\pi}j_{\pi}'}
[
[\tilde a_{j_{\nu}'} \times ( d_{\nu}^{\dagger} \times \tilde s_{\nu} )^{(2)}]^{(j_{\nu})}
\nonumber\\
&& \qquad \qquad \qquad \times
[(s_{\pi}^{\dagger} \times \tilde d_{\pi})^{(2)}\times \tilde a_{j_{\pi}'}]^{(j_{\pi})}]^{(\lambda)}.
\label{OBTD1}
\end{eqnarray}
For  reactions where the number of bosons is changed by one, and the core of the odd-odd nucleus
is added one neutron and removed one proton,
\begin{eqnarray}
&& {T}_{j_{\nu}j_{\pi}}^{(\lambda)} = 
\theta_{j_{\nu}}
\theta_{j_{\pi}}
[ ( s_\nu^{\dagger} \times \tilde a_{j_{\nu}})^{(j_{\nu})}
\times
(a_{j_{\pi}}^{\dagger} \times \tilde s_{\pi})^{(j_{\pi})}
]^{(\lambda)}
\nonumber\\
&& \quad 
+ \theta_{j_{\nu}}
\displaystyle \sum_{j_{\pi}'}
\theta_{j_{\pi}j_{\pi}'}
[ ( s_\nu^{\dagger} \times \tilde a_{j_{\nu}})^{(j_{\nu})}
\times
(a_{j_{\pi}'}^{\dagger} \times \tilde d_{\pi})^{(j_{\pi})}
]^{(\lambda)}
\nonumber\\
&& \quad + 
 \theta_{j_{\pi}} \displaystyle \sum_{j_{\nu}'}\theta_{j_{\nu}j_{\nu}'}
[ ( d_\nu^{\dagger} \times \tilde a_{j_{\nu}'})^{(j_{\nu})}
\times
(a_{j_{\pi}}^{\dagger} \times \tilde s_{\pi})^{(j_{\pi})}
]^{(\lambda)}
\nonumber\\
&& \quad + 
\displaystyle \sum_{j_{\nu}',j_{\pi}'}
\theta_{j_{\nu}j_{\nu}'}
\theta_{j_{\pi}j_{\pi}'}
[ ( d_\nu^{\dagger} \times \tilde a_{j_{\nu}'})^{(j_{\nu})}
\nonumber\\
&& \qquad \qquad \qquad \qquad
\times
(a_{j'_{\pi}}^{\dagger} \times \tilde d_{\pi})^{(j_{\pi})}]^{(\lambda)}
\label{OBTD2}
\end{eqnarray}

We proceed to compute  the matrix elements of the  operators of Eq.~\ref{OBTD1} and Eq.~\ref {OBTD2}.
For numerical calculations the $^{116}$Cd towards $^{116}$In we require to use the operator of Eq.~\ref{OBTD1}
and for the $^{116}$Cd toward $^{116}$Sn, we use the operator of Eq. \ref{OBTD2}. 

In the shell model, the eigenstates can be expressed in terms of the product of 
the proton and neutron spaces
\begin{equation}
|\Phi^{(A)}_{JM}\rangle = | \alpha_\pi J_\pi, \alpha_\nu J_\nu; JM \rangle
\end{equation}
The relationship between nuclear matrix elements of the shell model and the IBFFM is given by

\begin{equation}
\begin{array}{ccccc}
\langle \Phi^{(A')}_{J'}|
(a_{j_{\nu}}^{\dagger}
\times
\tilde a_{j_\pi})^{(\lambda)}
|\Phi^{(A)}_{J} \rangle
=
\langle \Psi^{(A')}_{J'}|
{T}_{j_{\nu}j_{\pi}}^{(\lambda)}
|\Psi^{(A)}_{J}\rangle,
\end{array}
\end{equation}

where $\Psi^{(A)}_{J}$ is the even-even nuclear wave function and 
$\Psi^{(A')}_{J'}$ the odd-odd nuclear wave function in the IBFFM which 
we will discuss in the next section.


\section{Description of odd-odd nuclei\label{sec4}}


In the IBFFM, odd-odd nuclei are described in terms of a system of 
$N_\pi$ proton bosons ($s_\pi$ and $d_\pi$) and $N_\nu$ neutron bosons 
($s_\nu$ and $d_\nu$) coupled to a single proton ($j_\pi$) and a single 
neutron ($j_\nu$). The proton and neutron orbitals are those of the 
active major shell for protons and neutrons, respectively. 

The Hamiltonian of the IBFFM for the odd-odd nuclei can be written as
\begin{eqnarray}
H=H^B+H^F_\pi+V^{BF}_\pi +H^F_\nu+V^{BF}_{\nu}+ V_{\rm res}
\label{Ham1}
\end{eqnarray}
$H^B$ is the IBM-2 Hamiltonian. $H^F_\pi$ and $H^F_\nu$ are the proton and 
neutron single-particle terms 
\begin{eqnarray}
H^F_\rho=\sum_{j_\rho} \epsilon_{j_\rho} \hat{n}_{j_\rho},
\end{eqnarray}
where $\epsilon_{j_\rho}$ is the quasi-particle energy of the extra nucleon, and $\hat{n}$ is the number operator. The quasi-particle 
energies $\epsilon_{j_\rho}$ are obtained within the BCS approximation with a gap $\Delta=12/\sqrt{A}$, where $A$ is the mass number of the 
even-even core nucleus. In the BCS, the quasi-particle energy are related to the single-particle level $\epsilon^{sp}$, the occupation 
probabilities $v_j$, and the fermi level $\lambda$ as follows: 
\begin{eqnarray}
\epsilon_j &=& \sqrt{(\epsilon^{sp}_j-\lambda)^2+\Delta^2}
\nonumber\\
v_j^2 &=& \frac{1}{2}\left( 1-\frac{\epsilon^{sp}_j-\lambda}{\epsilon_j}\right)
\nonumber\\
u_j^2 &=& 1-v^2_j 
\end{eqnarray}
$V^{BF}_{\pi}$ and $V^{BF}_{\nu}$ describe the 
core-particle coupling of the odd proton and odd neutron in the IBFM-2 model 
\cite{Alonso1984,Arias1985,Scholten1985} as the sum of a quadrupole term ($\Gamma_\rho$), an exchange term ($\Lambda_\rho$) and a monopole term ($A_\rho$)

\begin{equation}
V^{BF}_{\rho}= \Gamma_{\rho}  Q_{\rho'}^{(2)} \cdot  q_{\rho}^{(2)}
                   + \Lambda_{\rho} F_{\rho' \rho} 
                   + A_{\rho} \hat n_{d_{\rho'}} \cdot \hat n_{\rho}
\label{boson-fermion}
\end{equation}
where $\rho' \neq \rho$ and $\rho$, $\rho' =\nu$, $\pi$. The first term in 
Eq.~\ref{boson-fermion} is a quadrupole-quadrupole interaction with
\begin{eqnarray}
q^{(2)}_{\rho} &=& \sum_{j_\rho, j'_\rho} 
(u_{j_\rho} u_{j'_\rho} - v_{j_\rho} v_{j'_\rho})
Q_{j_\rho j'_\rho} (a^{\dagger}_{j_\rho} \times \tilde a_{j'_\rho})^{(2)}
\nonumber\\
Q^{(2)}_{\rho}&=& ( s_{\rho}^{\dagger} \times \tilde d_{\rho} 
+ d_{\rho}^{\dagger} \times \tilde s_{\rho} )^{(2)} 
+ \chi_{\rho} (d_{\rho}^{\dagger} \times \tilde d_{\rho})^{(2)}.
\end{eqnarray}
The second term is the exchange interaction 
\begin{eqnarray}
F_{\rho,\rho'} &=& -\sum_{j_\rho j'_\rho j''_\rho} \beta_{j_\rho j'_\rho} \beta_{j''_\rho j_\rho} \sqrt{\frac{10}{N_{\rho}(2j_\rho+1)}}
\nonumber\\
&& Q_{\rho'}^{(2)} \cdot :[(d_{\rho} \times \tilde a_{j''_\rho})^{(j_\rho)} \times 
(a^{\dagger}_{j'_\rho} \times \tilde s_{\rho})^{(j'_\rho)}]^{(2)}: + h.c. 
\end{eqnarray}
The coefficients $\beta_{j_\rho j'_\rho}$ are related to the single-particle matrix elements of the quadrupole operator $Q_{j_\rho j'_\rho}$ by
\begin{eqnarray}
\beta_{j_\rho j'_\rho} &=& (u_{j_\rho} v_{j'_\rho} + v_{j_\rho} u_{j'_\rho}) 
Q_{j_\rho j'_\rho} 
\nonumber\\
Q_{j_\rho j'_\rho} &=& \langle j_\rho \| Y^{(2)} \| j'_\rho \rangle.
\end{eqnarray}
The last term is the monopole-monopole interaction with 
\begin{eqnarray}
n_{d_\rho} &=& \sum_m d_{\rho,m}^{\dagger} d_{\rho,m} 
\nonumber\\
\hat n_\rho &=& \sum_{j_\rho} \hat{n}_{j_\rho} = 
\sum_{j_\rho,m} a^{\dagger}_{j_\rho,m} a_{j_\rho,m}
\end{eqnarray}

The residual interaction between the odd-proton and odd-neutron 
is defined as \cite{Brant1988,Yoshida2013} 
\begin{equation}
V_{\rm res}=H_{\delta}+H_{\sigma \sigma \delta}+H_{\sigma \sigma}+ H_{T},
\end{equation}
with 
\begin{eqnarray}
H_{\delta} &=& 4\pi V_{\delta}\delta(\vec r_{\pi}-\vec r_{\nu})\delta(r_{\pi}-R_{0})\delta(r_{\nu}-R_{0}),
\nonumber\\
H_{\sigma \sigma \delta} &=& 4\pi V_{\sigma \sigma \delta }
\delta(\vec r_{\pi}-\vec r_{\nu})
 (\vec \sigma_{\pi}\cdot \vec \sigma_{\nu})\nonumber\\
&& \times
\delta(r_{\pi}-R_{0})\delta(r_{\nu}-R_{0}),
\nonumber\\
H_{\sigma \sigma} &=& -\sqrt{3}V_{\sigma \sigma} \vec \sigma_{\pi}\cdot \vec \sigma_{\nu},
\nonumber\\
H_{T} &=& V_{T}\left[\frac{3(\vec \sigma_{\pi}\cdot \vec r_{\pi \nu})(\vec \sigma_{\nu}\cdot \vec r_{\pi \nu} )_{\pi\nu}}{r^{2}}-\vec \sigma_{\pi}\cdot \vec \sigma_{\nu}\right],
\end{eqnarray}
where $\vec r_{\pi \nu}= \vec r_{\pi}- \vec r_{\nu}$ and $R_{0}=1.2A^{1/3}$fm. 
The matrix elements of the residual interaction are calculated in the quasi-particle basis which is related to the particle basis by \cite{Yoshida2013}
\begin{eqnarray}
&& \langle j'_{\nu} j'_{\pi};J|V_{\rm res}| j_{\nu} j_{\pi};J \rangle_{qp} 
\nonumber\\
&& \qquad = (u_{j'_{\nu}} u_{j'_{\pi}} u_{j_{\nu}} u_{j_{\pi}}
           + v_{j'_{\nu}} v_{j'_{\pi}} v_{j_{\nu}} v_{j_{\pi}}) 
\nonumber\\
&& \qquad \qquad 
\times \langle j'_{\nu} j'_{\pi};J|V_{\rm res}|j_{\nu} j_{\pi};J \rangle 
\nonumber\\
&& \qquad \quad - (u_{j'_{\nu}} v_{j'_{\pi}} u_{j_{\nu}} v_{j_{\pi}}
            +v_{j'_{\nu}} u_{j'_{\pi}} v_{j_{\nu}} u_{j_{\pi}}) 
\nonumber\\
&& \qquad \qquad 
\sum_{J'}(2J'+1) \left\{ \begin{array}{ccc}
j'_{\nu} & j_{\pi} & J' \\
j_{\nu} & j'_{\pi} & J \end{array} \right\}
\nonumber\\
&& \qquad \qquad 
\times \langle j_{\nu}' j_{\pi};J'|V_{RES}|j_{\nu} j'_{\pi};J' \rangle,
\end{eqnarray}
where $v^2_{j}=1-u^2_j$ denotes the occupation probability.
 

\section{Numerical Results\label{sec5}}


\label{obtd}
\begin{table*}[htbp]
 \centering
 \caption{The IBM-2 parameters for the even-even nuclei taken from Ref.~\cite{Barea2013}}
 \begin{tabular}{ccccccccccccccccc}
 \hline
 \hline
 &&&$\epsilon_d$&$\kappa$&$\chi_\nu$&$\chi_\pi$&$\xi_1,\xi_3$&$\xi_2$&$c^{(0)}_\nu$&$c^{(2)}_\nu$&$c^{(4)}_\nu$
 &$v_{0}^{\rho}$&$v_{2}^{\rho}$\\
 Nucleus& $N_\nu$&$N_\pi$&(MeV)&(MeV)&(MeV)&(MeV)&(MeV)&(MeV)&(MeV)&(MeV)&(MeV)\\
 \hline
 $^{116}$Sn&8&0&1.32 &&&&&&-0.5& -0.22&-0.07&-0.06 &0.04 \\ 
 $^{116}$Cd &7&1&0.85 &-0.27&-0.58& 0.24& -0.18& -0.18& -0.15 & -0.06\\  
 \hline
 \end{tabular}
 \label{tab:par:ee}
\end{table*}
In the context of the IBFFM, the odd-odd nucleus is described by coupling a proton and a neutron to its core (even-even nucleus). 
Thus the first step is to construct the core nucleus. It is described by the IBM-2 Hamiltonian ($H^B$) \cite{Iachello1987}, which only 
depends on the neutron- and proton-boson degrees of freedom, the expression of the boson Hamiltonian $H^B$ is taken as in Ref.~\cite{Yoshida2013}. 

\begin{figure}[htbp]
 \centering
 \includegraphics[scale=0.7]{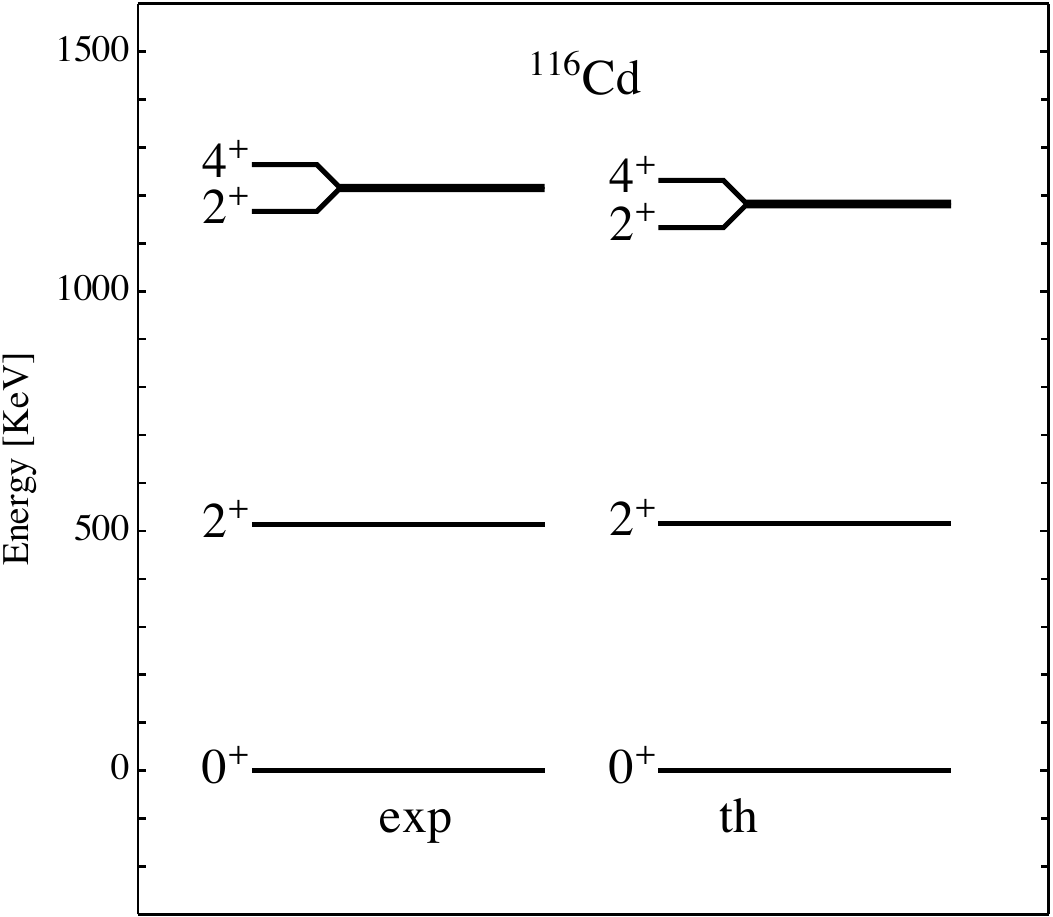} 
 \caption{The energy levels obtained in the calculation in comparison with the available experimental data for even-even nucleus $^{116}$Cd \cite{Blachot2010}.}
 \label{fig:116Cd}
\end{figure}

To construct the odd-odd wave function of $^{116}$In, we consider $^{116}$Cd as the core nucleus.
The parameters for the core nucleus are taken from the literature \cite{Barea2013}. They are shown in Table \ref{tab:par:ee}.
The spectrum of $^{116}$Cd obtained in our calculation is presented in 
Fig.~\ref{fig:116Cd} compared to the available experimental data \cite{Blachot2010}. 
We can see that the agreement is quite good \cite{Barea2009}.

\begin{figure}[htbp]
 \centering
 \includegraphics[scale=0.7]{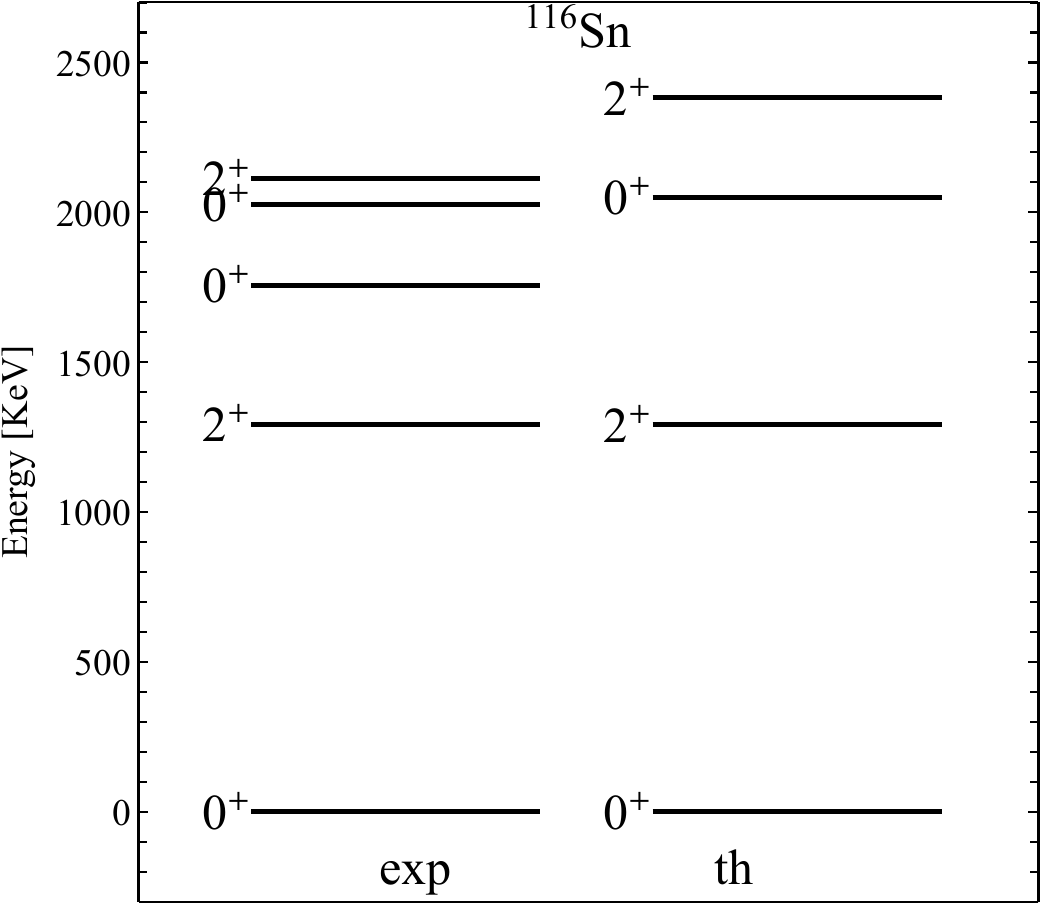} \caption{The spectrum of the $^{116}$Sn in comparison with the experimental data \cite{Blachot2010}.}
 \label{fig:116Sn}
\end{figure}

The second step is to construct the two odd-even associated nuclei, i.e. the core nucleus plus an extra neutron and the core nucleus plus an 
extra proton. It allows us to get a reliable set of parameters that we use to construct the odd-odd wave function. In this case, the odd-even 
nuclei are described in the context of the IBFM-2 \cite{Alonso1984,Arias1985,Iachello1991}  where the degrees of freedom of the extra nucleon are taken into account. 
In the IBFM-2, the Hamiltonian is given by 
\begin{eqnarray}
H=H^B+H^F_\rho+V^{BF}_{\rho}, 
\label{hibfm2}
\end{eqnarray}
where $H^B$ is the boson Hamiltonian that describes the core nucleus, the label $\rho$ refers to the $\pi$ (extra proton) or $\nu$ (extra neutron) 
is added in the even-even core to form the odd-even nucleus. 

 The single-particle energies $\epsilon^{sp}$   are obtained by solving the Schr\"odinger equation  with the phenomenological   Woods-Saxon potential based upon the sum of a spin-independent central potential, a spin-orbit potential, and the Coulomb potential\cite{brown2005lecture}, 
 \begin{equation}
 V_{ws}(r)=V_{o}(r)+V_{so}(r) \vec l \cdot  \vec s+ V_{c}(r),
 \end{equation}
where $V_{o}$(r) is the spin-independent central potential:
\begin{equation}
V_{o}(r)= V_{o}f_{o}(r),
\end{equation}
with a fermi shape
\begin{equation}
f_{o}(r)=\frac{1}{1+[\text{exp}(r-R_{0})/a_{0}]},
\end{equation}
$V_{so}(r)$ is the spin-orbit potential:
\begin{equation}
V_{so}(r)=V_{so}\frac{1}{r}\frac{df_{so}(r)}{dr},
\end{equation}
with
\begin{equation}
f_{so}(r)=\frac{1}{1+[\text{exp}(r-R_{so})/a_{0}]},
\end{equation}
and $V_{c}(r)$ is the Coulomb potential for protons based upon the Coulomb potential of a
sphere of radius $ R_{c}$:
\begin{equation}
V_{c}(r)= \frac{Z e^{2}}{r} \text{ for } r \geqslant R_{c},
\end{equation}
and
\begin{equation}
V_{c}(r)= \frac{Z e^{2}}{R_{c}}\left[\frac{3}{2}- \frac{r^{2}}{2R_{c}^{2}},
\right] \text{ for }r\leqslant R_{c}.
\end{equation}
The radii $R_{0},R_{so} $, and $ R_{c}$ are usually expressed as:
\begin{equation}
R_{i}=r_{i}A^{1/3}.
\end{equation}
since the average proton-neutron potential is stronger than the average neutron-neutron (or proton-proton) potential, thus in a nuclei with a neutron excess, the protons will feel a stronger potential than the neutrons, for this reason we take
\begin{equation}
V_{op}=V_{0}+\frac{N-Z}{A}V_{1} \text{ for protons},
\end{equation}
and
\begin{equation}
V_{on}=V_{0}-\frac{N-Z}{A}V_{1} \text{ for neutrons}.
\end{equation}

The single particle energies  are obtained  by solving the  Woods-Saxon with a typical set of parameters   with  $V_{0} = -53$ MeV, $V_{1} = -30$ MeV and 
$V_{\rm so} =22$ MeV for the strengths, and
$r_{0} =r_{\rm so} =1.29$ fm and 
$a_{0} =a_{\rm so} =0.70$ fm for the geometry. 
For the Coulomb term the radius is a little smaller with $r_{c} = 1.20$ fm. 
The quasi-particle energies and occupation probabilities were obtained solving the BCS equations with the single particle energies calculated and reported in 
the Table \ref{tab:spe}. 
The value of $\lambda$ is constrained to the conservation of the number of particles as follows:
\begin{eqnarray}
2N_\rho=\sum_{j_\rho} v^2_{j_\rho}(2j_\rho+1). 
\end{eqnarray} 

\begin{table}[ht]
\label{tablespe1}
\caption{The single particle energies, quasi-particle energies
and occupation probabilities used in the OBTD for
$116$Sn to $116$In. For the neutron in the 50-82 shell as a particle
and for the proton 28-50 shell as a hole. }
 \begin{tabular}{cc}
 Neutron case \\
 \end{tabular}
 \\
\begin{tabular}{ccccc}
 \hline
orbit& spe. (MeV) &qspe.(MeV)\\ 
n l j& $\epsilon_{j}^{sp}$&$\epsilon_{j}^{qpe}$&$v_{j}^{2}$\\ 
 \hline
2d5/2& 2.7090& 2.1439 & 0.0735 \\
1g7/2 &1.5300 & 1.2940 & 0.2489\\
3s1/2& 0.3730 & 1.2286 & 0.7064 \\
1h11/2&0.0000 & 1.4237 & 0.8091 \\
2d 3/2&-0.2890 & 1.6184 & 0.8612 \\
\end{tabular}
\\
 \begin{tabular}{cc}
 Proton case \\
 \end{tabular}
 \\
\begin{tabular}{ccccc}
 \hline
orbit& spe. (MeV) &qspe.(MeV)\\ 
n l j& $\epsilon_{j}^{sp}$&$\epsilon_{j}^{qp}$&$v_{j}^{2}$\\ 
 \hline
2p1/2 &0.2610& 3.8569 & 0.9789 \\
2p3/2& 1.8900 & 2.3440 & 0.9405\\
1f5/2&2.3530 & 1.9485 & 0.9110 \\
1g9/2& 0.0000 & 4.1075 & 0.9814 \\
\end{tabular}
 \label{tab:spe}
\end{table}

The odd-odd $^{116}$In nucleus uses the quasi-particle energies 
obtained from the single-particle energies for the odd-neutron nucleus 
$^{115}$Cd and the odd-proton nucleus $^{117}$In. Both odd-even nuclei 
have the same even-even core nucleus $^{116}$Cd. 
The core-particle coupling parameters reported in Table \ref{tab:par:oe} 
were obtained by fitting the experimental data for $^{115}$Cd and $^{117}$In. 

\begin{table}[htbp]
\centering
\caption{Boson-fermion interaction parameters for the odd-even nuclei associate to $^{116}$Cd core.}
\begin{tabular}{cccc}  
\hline
&$\Gamma_\rho$&$\Lambda_\rho$&$A_\rho$\\
&(MeV)&(MeV)&(MeV)\\
\hline
proton in $^{117}$In &-0.5&4.0 &-1.0\\
neutron in $^{115}$Cd &2.49&2.5 &1.0\\
\hline
\end{tabular}
\label{tab:par:oe}
\end{table}

For the odd-odd nucleus $^{116}$In, the tensor and Surface Delta Interaction (SDI) play an essential role. 
In this work, the parameters are obtained by minimization of the Mean Square Error (MSE) on the test set given by
$\text{MSE}=\frac{1}{N}\sum_n^N(f_n(x)-c_n)^2$ where $f_n(x)$ are our theoretical calculations and $c_n$ are the experimental values \cite{Blachot2010}.
We are interested in studying the low-lying spectra of the odd-odd nucleus.
We consider the first three levels of the $^{116}$In, and this procedure may be applied to many levels as are needed. It is computed the Mean Square Error for different values of the parameter of Tensor and SDI.
We have seen that the MSE is very sensitive to the parameters, as shown in 
Fig.~\ref{map2}.

\begin{figure}[htbp]
\centering
\includegraphics[scale=0.6]{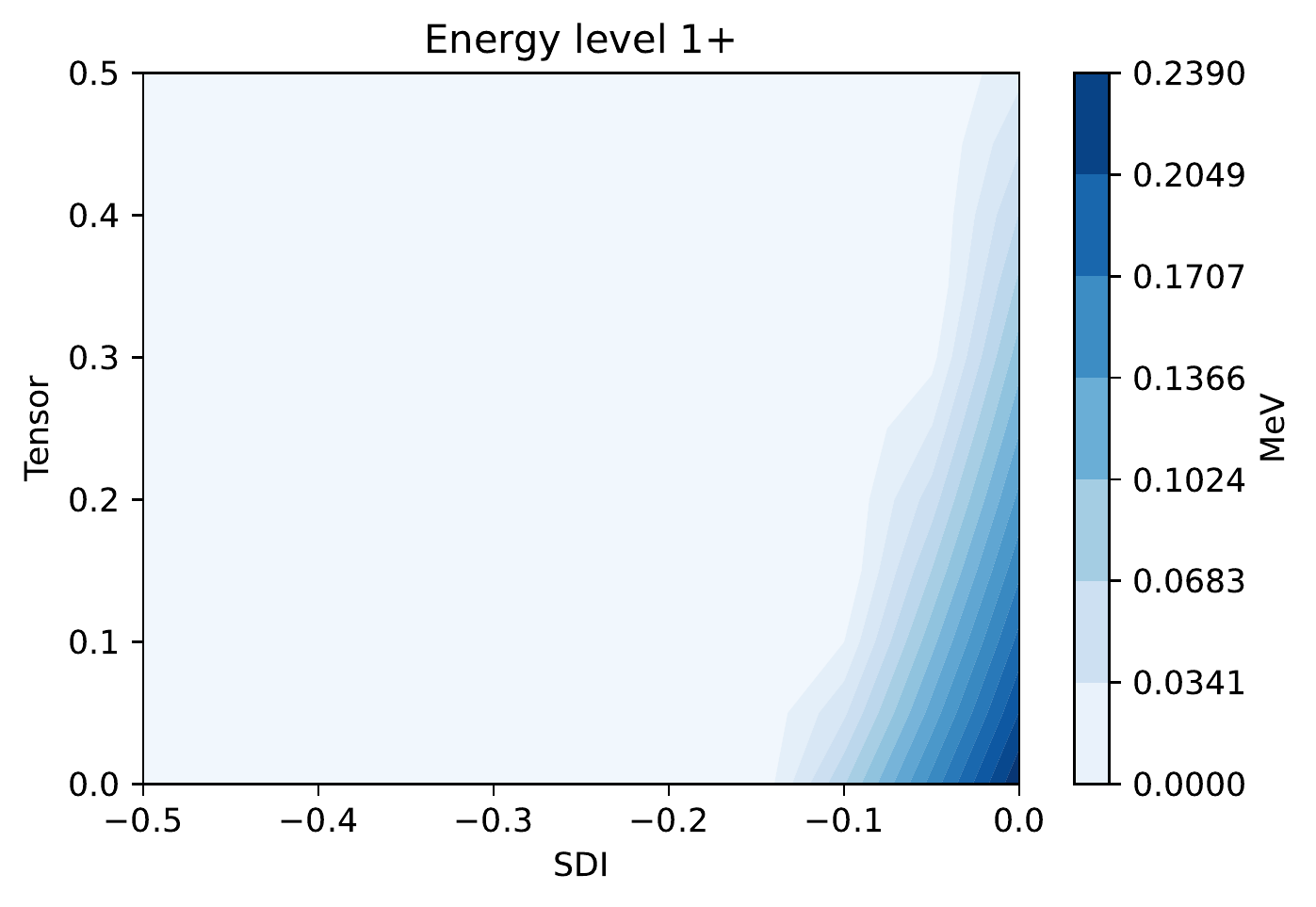}
 \caption{Energy Surface $^{116}$In.}
 \label{ground}
\end{figure}
 
The theoretical energy surface of the ground state $1+$ of $^{116}$In is depicted in the Fig. \ref{ground}.
We found that the MSE is very sensitive to the parameters, as we see in Fig.~\ref{map2}.
\begin{figure}[htbp]
 \centering
 \includegraphics[scale=0.6]{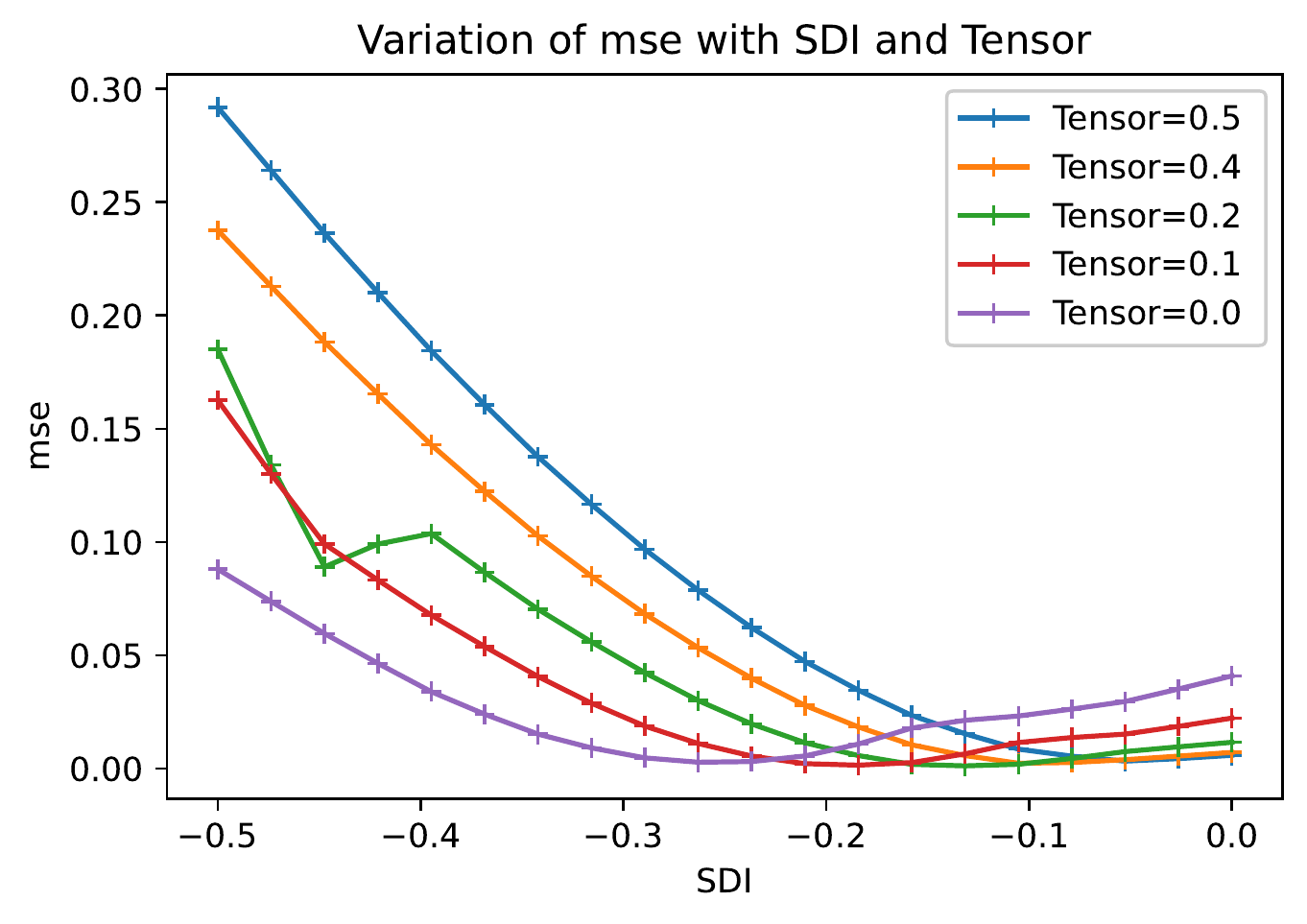} 
 \caption{
 Values of the Mean Square Error (MSE) for different values of the parameter SDI and Tensor for the best fit $^{116}$In.}
\label{map2}
\end{figure}

The minimization procedure consists of the creation of an
uninterpolated contour surface with a given set of parameters
over a grid of dimensions at least of dimension 11$\times$11.
 After that, we generate an interpolated function of a grid dimension 30$\times$30, which is used to determine 
 the best parameters to fit the experimental data.

\begin{figure}[htbp]
 \centering
 \includegraphics[scale=0.34]{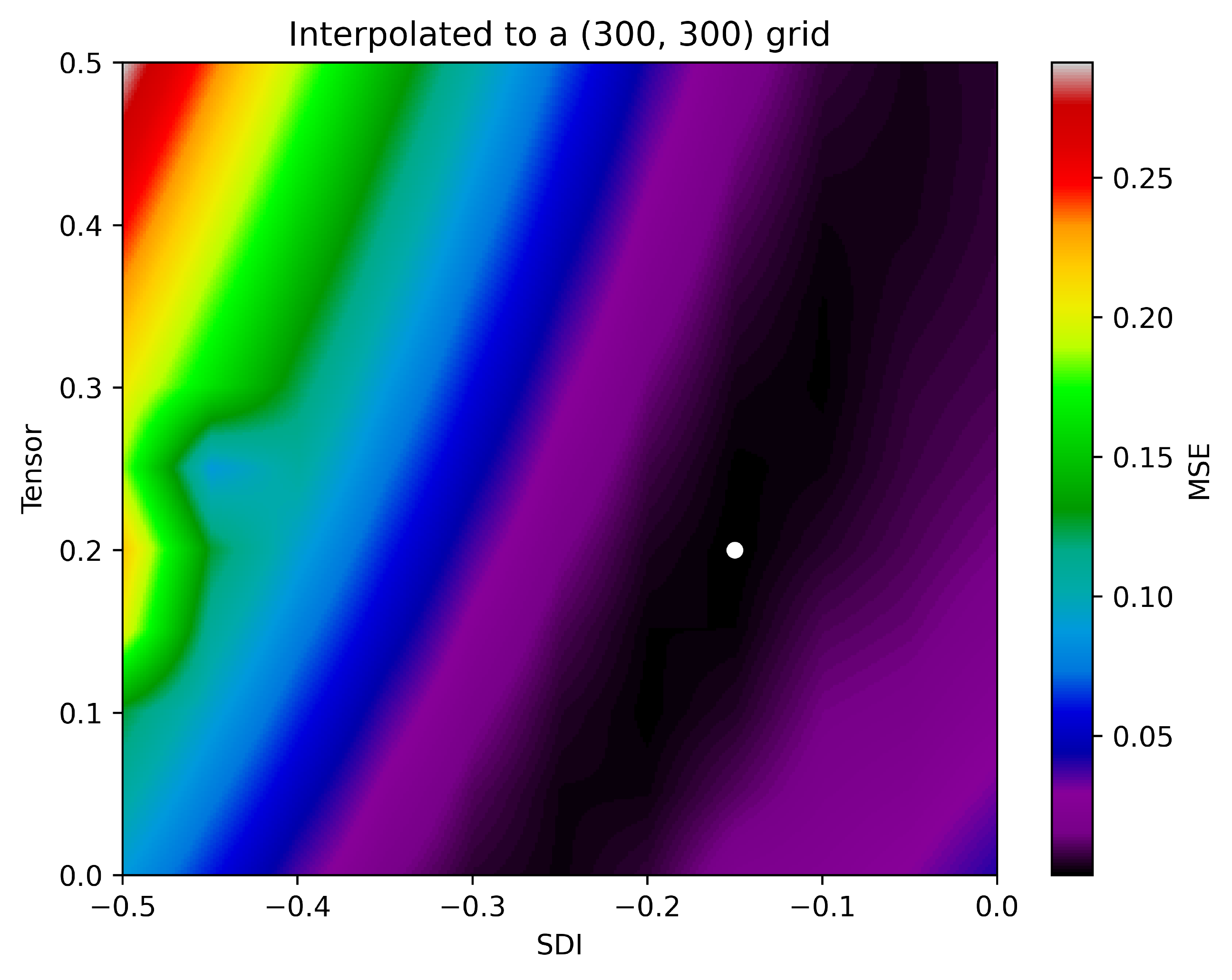} 
 \caption{Interpolated surface of the Mean Square Error. The minimum of the surface is shown with the white dot.}
 \label{fit1}
\end{figure}

\begin{figure}[htbp]
 \centering
 \includegraphics[scale=0.8]{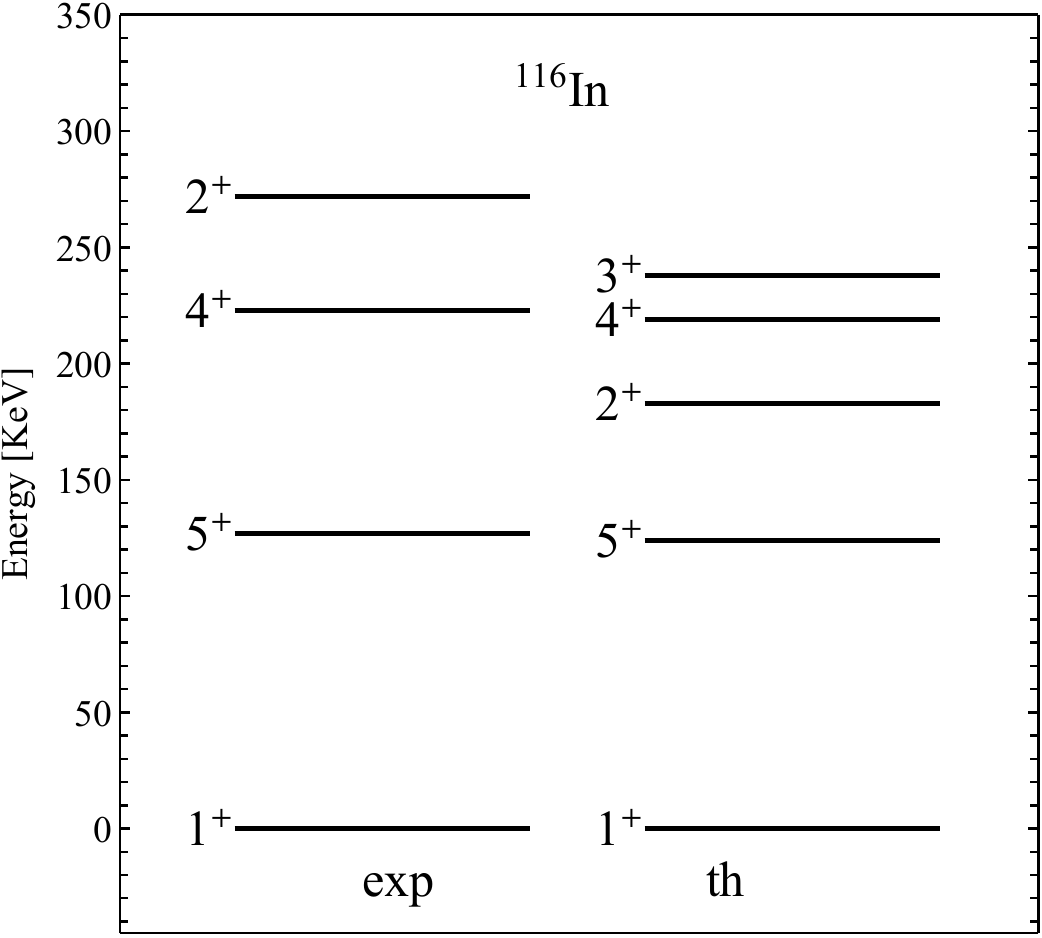} 
 \caption{The spectrum of $^{116}$In.}
 \label{fig:116In}
\end{figure}

The fitting method is given by the Nelder-Mead simplex algorithm to find the minimum of a function of one or more variables.
The library used is the SciPy that uses NumPy, the Machine Learning Python library
used for scientific and technical computing.
We obtain the minimum of the surface at the point with SDI -0.1500 and Tensor 0.200. with min error 6.53e-05.
The result is depicted in the Fig \ref{fit1}. 
The spectrum of the $^{116}$In is presented in the Fig. \ref{fig:116In}.
 \begin{table}
 \centering
  \caption{ The parameters of the proton-neutron residual interaction.}
 \begin{tabular}{cccccccc}  
 \hline
 Parameter && Value (in MeV) \\
 &&$^{116}$In&\\
 \hline
 $V_\delta$&& -0.1500.&\\
 $V_{\sigma\sigma}$&&0.0&\\
 $V_T$&&0.200&\\
 \hline  
 \end{tabular}
 \label{tab:booktabs}
\end{table}
This method also can be used for the fitting parameters for the odd-even nuclei, and even-even nuclei. 
 \begin{table}[htbp]
 \centering
 \caption{Theoretical energy Levels obtained in the calculation in comparison with the available experimental data for odd-odd 
 nucleus $^{116}$In \cite{Blachot2010}.}
 \begin{tabular}{cccc} 
 \hline
 &$^{116}$In&&\\
 \hline
 Spin &Spin& E[KeV]& E[KeV] \\
 Exp. & Th. &Exp.&Cal. \\
  \hline
$1^+$ &$1^+$ &0.0  &0.0 \\
$5^+$ &$5^+$ &127.267 &120.4 \\
$4^+$ &$4^+$ &223.330 &219.8 \\  
$2^+$ &$2^+$ &272.966 &183.4 \\  
$4^+,5^+$&$4^+$ &313.476 &297.5 \\  
$4^+$ &$4^+$ &425.930 &468.1 \\  
$4^+,5^+$&$5^+$ &460.0 &336.6 \\  
$3^+$ &$3^+$ &508.241 &238.2 \\  
 \\
 \hline

 \end{tabular}
 \label{tab:energylevelIn}
\end{table} 

In this work, the parameters are fitted to reproduce the experimental level energies
and depicted in the Figs.~\ref{fig:116In} and are reported in Table~\ref{tab:energylevelIn}.

In the following section, we compute the spectroscopic amplitudes by using the previous nuclear wave functions.

\subsection{Spectroscopic Amplitudes in IBFFM}

The model space of transition is given by the model space of the odd-odd nucleus.  
The model is given by five active neutron orbitals $2d5/2$, $1g7/2$, $3s1/2$, $2d3/2$ and $1h11/2$, and by four active proton orbitals $2p1/2$, $2p3/2$, $1f5/2$ and $1g9/2$. 

\begin{figure}[htbp]
\centering
 \includegraphics[scale=0.35]{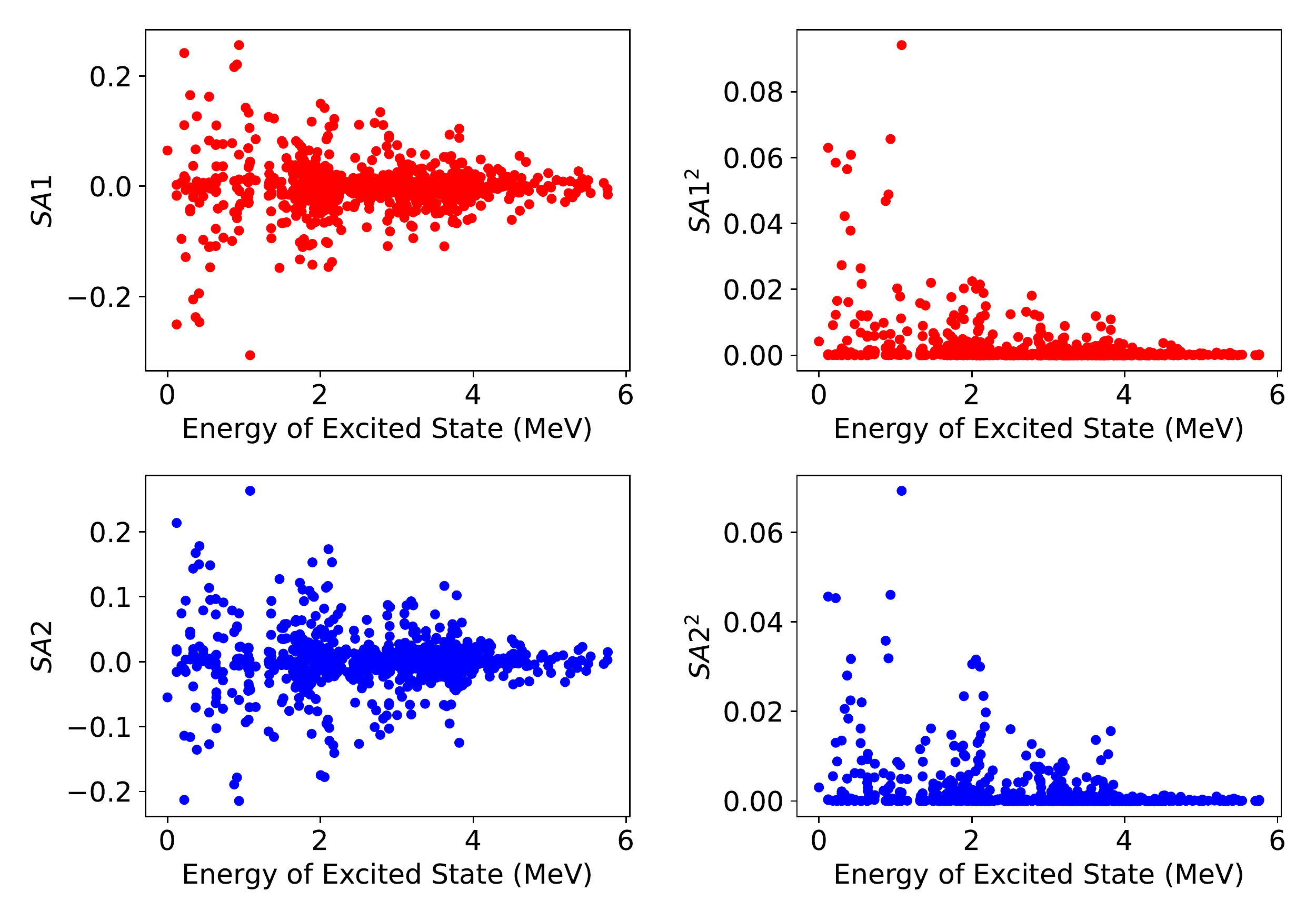} 
 \caption{Spectroscopic Amplitudes for the transition $^{116}$Cd $\rightarrow$ $^{116}$In denoted as SA1
 and transitions $^{116}$Sn $\rightarrow$ $^{116}$In denoted as SA2. The squared of the spectroscopic amplitudes are denoted as SA1$^2$and 
 SA2$^2$ respectively.}
 \label{fig7}
\end{figure}

The  nuclear states  of  the $^{116}$In system are in our model restricted to be
$J_P=0^+$, $1^+, 2^+, 3^+, 4^+, 5^+, 6^+, 7^+, 8^+$, and $9^+$, within the first 40 excited states.
The calculated excitation energies (in MeV). We have decided to use the calculated rather than experimental energies because some
of the experimentally known states may not correspond.
The odd-odd nuclei's parities positives are chosen because we are interested in the odd-odd nucleus's ground state.
This fact affects the selection rules of the available orbitals of the transition operator for neutrons and protons.
The neutron and proton model space chosen in the IBFFM depends on the active protons and neutrons' shell.
The neutron's shell chosen for the $^{116}$In is the 50-82 shell, and for the protons, the 28-50 shell.
For the numerical calculations of the normalization coefficients of the single nucleon transfer of neutron and proton for particle and hole coupling, we have considered the first 40 excited states.
The numerical results of the spectroscopic amplitudes are exhibited in the Table \ref{result2} for various cases of interest. 
We denote the spectroscopic amplitudes as SA1 the transitions $^{116}$Cd to $^{116}$In
and SA2 for the transitions $^{116}$Sn to $^{116}$In.
It was found that the squared
spectroscopic amplitudes for both transitions
(SA1$^2$ and SA2$^2$) tends to zero within higher energies 6 MeV (see Fig. \ref{fig7} ).
That means that the main contributions of the intermediate states are
$\le 6$ MeV. Therefore we do not need an infinite number of intermediate states to compute the SA.

\begin{table}[ht]
\caption{Spectrocopic Amplitudes SA (fifth column) and One Body Transition Densities OBTD (last column) calculated using IBFFM, for the 
ground state $^{116}$Sn $0^{+}$ and $^{116}$Cd $0^{+}$ to the the firsts states of $^{116}$In, and the third column and fourth column are 
listed the active orbitals for neutrons and protons $(j_{\nu}, j_{\pi})$ respectively.}
\begin{tabular}{llllrr}
\toprule
\hline
Initial nucleus & Final nucleus & $j_\nu$ & $j_\pi$ &SA &OBTD \\
\hline
\midrule
 116Sn( $0_1^+$) & 116In( $1^+_1$) &  1g7/2 &  1g9/2 & -0.05495 & -0.0317 \\
    & 116In( $5^+_1$) &  2d5/2 &  1g9/2 & 0.01913 & 0.0058 \\
    & 116In( $5^+_1$) &  1g7/2 &  1g9/2 & 0.01544 & 0.0047 \\
    & 116In( $5^+_1$) &  3s1/2 &  1g9/2 & 0.21360 & 0.0644 \\
    & 116In( $5^+_1$) &  2d3/2 &  1g9/2 & -0.01582 & -0.0048 \\
    & 116In( $4^+_1$) &  2d5/2 &  1g9/2 & -0.01154 & -0.0038 \\
    & 116In( $4^+_1$) &  1g7/2 &  1g9/2 & -0.01244 & -0.0041 \\
    & 116In( $4^+_1$) &  3s1/2 &  1g9/2 & -0.21281 & -0.0709 \\
    & 116In( $4^+_1$) &  2d3/2 &  1g9/2 & -0.11402 & -0.0380 \\
    & 116In( $2^+_1$) &  2d5/2 &  1g9/2 & -0.00634 & -0.0028 \\
    & 116In( $2^+_1$) &  1g7/2 &  1g9/2 & 0.07435 & 0.0333 \\
    & 116In( $4^+_2$) &  2d5/2 &  1g9/2 & 0.00338 & 0.0011 \\
    & 116In( $4^+_2$) &  1g7/2 &  1g9/2 & -0.11610 & -0.0387 \\
    & 116In( $4^+_2$) &  3s1/2 &  1g9/2 & 0.04128 & 0.0138 \\
    & 116In( $4^+_2$) &  2d3/2 &  1g9/2 & 0.04624 & 0.0154 \\
    & 116In( $3^+_1$) &  2d5/2 &  1g9/2 & 0.00319 & 0.0012 \\
    & 116In( $3^+_1$) &  1g7/2 &  1g9/2 & 0.09394 & 0.0355 \\
    & 116In( $3^+_1$) &  2d3/2 &  1g9/2 & -0.01571 & -0.0059 \\
 116Cd( $0_1^+$) & 116In( $1^+_1$) &  1g7/2 &  1g9/2 & 0.06484 & 0.0374 \\
    & 116In( $5^+_1$) &  2d5/2 &  1g9/2 & -0.01732 & -0.0052 \\
    & 116In( $5^+_1$) &  1g7/2 &  1g9/2 & -0.01628 & -0.0049 \\
    & 116In( $5^+_1$) &  3s1/2 &  1g9/2 & -0.25093 & -0.0757 \\
    & 116In( $5^+_1$) &  2d3/2 &  1g9/2 & 0.00255 & 0.0008 \\
    & 116In( $4^+_1$) &  2d5/2 &  1g9/2 & 0.01031 & 0.0034 \\
    & 116In( $4^+_1$) &  1g7/2 &  1g9/2 & 0.01832 & 0.0061 \\
    & 116In( $4^+_1$) &  3s1/2 &  1g9/2 & 0.24176 & 0.0806 \\
    & 116In( $4^+_1$) &  2d3/2 &  1g9/2 & 0.11084 & 0.0369 \\
    & 116In( $2^+_1$) &  2d5/2 &  1g9/2 & 0.00679 & 0.0030 \\
    & 116In( $2^+_1$) &  1g7/2 &  1g9/2 & -0.09558 & -0.0427 \\
    & 116In( $4^+_2$) &  2d5/2 &  1g9/2 & -0.00186 & -0.0006 \\
    & 116In( $4^+_2$) &  1g7/2 &  1g9/2 & 0.16537 & 0.0551 \\
    & 116In( $4^+_2$) &  3s1/2 &  1g9/2 & -0.04115 & -0.0137 \\
    & 116In( $4^+_2$) &  2d3/2 &  1g9/2 & -0.04573 & -0.0152 \\
    & 116In( $3^+_1$) &  2d5/2 &  1g9/2 & -0.00669 & -0.0025 \\
    & 116In( $3^+_1$) &  1g7/2 &  1g9/2 & -0.12854 & -0.0486 \\
    & 116In( $3^+_1$) &  2d3/2 &  1g9/2 & 0.01363 & 0.0052 \\
\bottomrule
    \hline
\end{tabular}
\label{result2}
\end{table} 


\section{Summary and concluding remarks\label{sec6}}

It was presented for the first time the explicit operator of the IBFFM for calculating one body transition densities in terms of active orbitals.
It was derived the odd-even and odd-odd parameters by using an implemented method of fitting using Machine learning libraries, 
which allowed computing more realistic nuclear wave functions and presented the spectroscopic amplitudes needed to calculate the double charge exchange without closure. 
For spins of intermediate states higher than 6 MeV, the SA tends to zero.
To summarize, the spectroscopic amplitudes of the $^{116}$Cd, $^{116}$Sn to $^{116}$In nuclei are investigated within the IBFFM approach. The even-even boson-core 
Hamiltonian, and essential building blocks of the particle-boson coupling Hamiltonians, i.e., single-particle energies and odd particles' occupation probabilities.  A few coupling constants for the boson-fermion Hamiltonians and the residual neutron-proton interaction remain the only free parameters of the model. 
They are determined to reasonably reproduce the low-energy levels of each of the neighboring odd-$A$ and odd-odd nuclei.
The IBM and IBFFM wave functions obtained after diagonalization 
of the corresponding Hamiltonians for the parent and daughter nuclei 
For the first time, the parameters of the odd-odd $^{116}$In are presented.
This is attributed to the combination of various factors adopted in the theoretical procedure, such as the chosen boson-fermion coupling constants, 
residual neutron-proton interaction, and underlying microscopic inputs.
The spectroscopic amplitudes converge to zero after a certain energetic value of the odd-odd nuclei's intermediate states, as shown in Fig \ref{fig7}. It has been correlations of SA between the case of the same core and different core.
\acknowledgments{}

We acknowledge fruitful discussion with Prof. Dr. Horst Lenske   and Dr. Maria Colonna. 
This work was supported in part by the INFN Sezione di Genova, and in part by grant IN101320 from DGAPA-UNAM, Mexico. 

\bibliographystyle{apsrev4-1}
\bibliography{refs3}
\end{document}